# Gambling scores in earthquake prediction analysis

G. Molchan and L. Romashkova

*International Institute of Earthquake Prediction Theory and
Mathematical Geophysics, Russian Academy of Science, Moscow, Russia.
The Abdus Salam International Centre for Theoretical Physics,
SAND Group, Trieste, Italy.
E-mail: molchan@mitp.ru, lina@mitp.ru*

**Abstract**
The number of successes $\nu^+$ and the normalized measure of space-time alarm $\tau$ are commonly used to characterize the strength of an earthquake prediction method and the significance of prediction results. To evaluate better the forecaster's skill, it has been recently suggested to use a new characteristic, the gambling score $R$, which incorporates the difficulty of guessing each target event by using different weights for different alarms. We expand the class of $R$-characteristics and apply these to the analysis of results of the M8 prediction algorithm. We show that the level of significance $\alpha$ strongly depends (1) on the choice of weighting alarm parameters, (2) on the partitioning of the entire alarm volume into component parts, and (3) on the accuracy of the spatial rate of target events, $\lambda(dg)$. These tools are at the disposal of the researcher and can affect the significance estimate in either direction. All the $R$-statistics discussed here corroborate that the prediction of $8.0 \leq M < 8.5$ events by the M8 method is nontrivial. However, conclusions based on traditional characteristics $(\nu^+, \tau)$ are more reliable owing to two circumstances: $\tau$ is stable since it is based on relative values of $\lambda(\cdot)$, and the $\nu^+$ statistic enables constructing an upper estimate of $\alpha$ taking into account the uncertainty of $\lambda(\cdot)$.

## 1. Introduction

Earthquake prediction of the yes/no type usually deals with two characteristics: the rate of failures-to-predict, $n$, and the normalized measure of space-time alarm, $\tau$ (see, e.g., *Molchan, 2003*). In these terms one can characterize the strength of a prediction method and the significance of prediction results. The $(n, \tau)$ vector as an integral characteristic may mask inefficiency of a method in some parts of the space-time monitoring area. This is somewhat compensated by relaxed requirements on the accuracy of the spatial rate of target events, $\lambda(dg)$, as well as by the ease with which the uncertainty of $\lambda(dg)$ can be incorporated in estimating the significance (*Molchan, 2010; Molchan & Romashkova, 2010*).

The problem of choosing a suitable quantity to characterize the strength of a prediction method, for instance, in the atmospheric sciences, is treated in an extensive literature (see, e.g., the review by *Joliffe & Stephenson, 2003*). Recently *Zhuang (2010)* and *Zechard & Zhuang (2010)* suggested the gambling score (GS) approach to earthquake prediction analysis in which different alarms are weighted differently; for example, a successful alarm with smaller prior probability of a target event is considered as more valuable. In this case the role of local accuracy of $\lambda(dg)$ in prediction analysis becomes significantly greater. In addition, the weighting of alarms may be done in different ways. For these reasons the GS approach, its motivation, and practical applications should be discussed.

In this study we generalize the GS approach and discuss its applications to some earthquake prediction techniques. These are the M8 algorithm by *Keilis-Borok & Kossobokov (1990)*, which is designed for the prediction of $M \geq 8.0$ events worldwide, and the Reverse Tracing of Precursors (RTP) method by *Keilis-Borok et al. (2004)*. The M8 algorithm was examined recently by these authors (*Molchan & Romashkova, 2010*) and the RTP technique by *Zechar & Zhuang (2010)*. This considerably simplifies our task of a preliminary analysis of data and predictions, so that we can concentrate on the methodology of the GS approach.

## 2. The Gambling Score approach
### 2.1. The R-Score

Let $Y = (y_1, \ldots, y_{N_a})$ be a random binary sequence with the probabilities of outcome for the i-th component equal to
$$P(y_i = 1) = p_i, \ P(y_i = 0) = 1 - p_i. \tag{1}$$
Here, $y_i = 1$ may be interpreted as the occurrence of a target event in a space-time domain $A_i = G_i \times T_i$, and $y_i = 0$ as no event in $A_i$.

Let $X = (x_1, \ldots, x_{N_a})$ be another binary vector to be treated as a prediction of $Y$: $x_i = 1$ means an alarm, i.e., the *positive prediction* of a target event in $A_i$, while $x_i = 0$ means no alarm, i.e., the *negative prediction* of a target event in $A_i$.

To be able to specify the quality of a prediction or the proximity of $X$ and $Y$, an appropriate measure (score) $R(X, Y)$ is used. The choice of $R$ may depend on a variety of factors like scientific and applied goals in the study of a prediction method, requirements on the stability of the distribution of $R$ given a reference seismicity model, etc.

The GS approach is of interest as being a tool for constructing meaningful models of $R$. This approach treats $R$ as the net gain of the forecaster in a sequence of $N_a$ trials. In the i-th trial the forecaster may stake $b_+(p_i)$ on the outcome $y_i = 1$ or $b_-(p_i)$ on the outcome $y_i = 0$. In case of success the forecaster gets $a_+(p_i)$ or $a_-(p_i)$, respectively. The net gain is

$$R(X, Y) = \sum_{i=1}^{N_a} [a_+(p_i) x_i y_i - b_+(p_i) x_i \bar{y}_i - b_-(p_i) \bar{x}_i y_i + a_-(p_i) \bar{x}_i \bar{y}_i], \tag{2}$$

where we denoted $\bar{c} = 1 - c$. It is natural to assume that the forecaster's expected gain under (1) in each trial is zero, i.e.,
$$E_Y [a_+(p_i) y_i - b_+(p_i) \bar{y}_i] = 0 = E_Y [a_-(p_i) \bar{y}_i - b_-(p_i) y_i],$$
where $E_Y$ denotes the expectation with respect to the distribution of $Y$. This assumption restricts the number of unknown functions $\{a_\pm, b_\pm\}$ to two, because it gives
$$a_+(p)/b_+(p) = \bar{p}/p, \ a_-(p)/b_-(p) = p/\bar{p}. \tag{3}$$

As a result, $R$ transforms to
$$R(X, Y) = \sum_{i \geq 1} w(p_i)(x_i - p_i)(y_i - p_i) + \sum_{i \geq 1} v(p_i)(y_i - p_i),$$
where
$$w(p) = b_+(p)/p + b_-(p)/\bar{p}, \ v(p) = b_+(p) - b_-(p).$$

In order to further restrict the number of unknown functions, we shall interpret $a_+(p)$ and $a_-(p)$ as one and the same measure of the complexity in guessing the random events $\{y = 1\}$ and $\{y = 0\}$, respectively. If the measure depends on the probability of the event only, then according to (1) we have

$$a_-(p) = a_+(1-p). \tag{4}$$

By (3, 4), we get an analogous relation for $b_+$ and $b_-$:

$$b_-(p) = b_+(1-p). \tag{5}$$

The corollary is that $w(p)$ is symmetric: $w(p) = w(1-p)$, while $v$ is anti-symmetric: $v(p) = -v(1-p)$.

When dealing with the prediction of large events, the typical situation involves a small $p$, $p < 1/2$. In order to stimulate the prediction of rare events ($p < 1/2$), the natural requirement is $b_+(p) \leq b_-(p)$ for the stakes and $a_+(p) \geq a_-(p)$ for the gains. The simplest choice

$$b_+(p) = b_-(p), \quad 0 < p \leq 1/2 \tag{6}$$

is sufficient to satisfy the requirement on gains. By (3, 6), one has

$$a_+(p)/a_-(p) = (p^{-1} - 1)^2 \geq 1, \quad p \leq 1/2.$$

If $a_+(p)$ as a measure of complexity decreases on (0, 1), it follows that the requirement on gains is satisfied without (6): $a_+(p) \geq a_+(1-p) = a_-(p)$ when $p \leq 1/2$.

The requirement (6) is convenient, because $v(p) = 0$, so $R$ is

$$R(X,Y) = \sum_{i \geq 1} w_i (x_i - p_i)(y_i - p_i), \quad w_i = w(p_i), \tag{7}$$

where $w(p)$ specifies all basis functions $\{a_\pm, b_\pm\}$:

$$w(p) = b_\pm(p)[p(1-p)]^{-1} = a_+(p)(1-p)^{-2} = a_-(p)p^{-2}. \tag{8}$$

*Zechar & Zhuang (2010)* use the score (2, 3) with $b_\pm(p) = 1$. In virtue of (8) the score is consistent with (7) when

$$w(p) = [p(1-p)]^{-1}. \tag{9}$$

The representation (7) is of interest since it can be treated in purely geometric terms, i.e., as a weighted correlation of the vectors $X - P$ and $Y - P$, where $P = (p_1, \ldots, p_{N_a})$ is the mean of vector $Y$ under (1). The weight $w_i$ may reflect the degree of importance of the i-th prediction or the complexity involved in guessing the result $y_i = 1$ when $p \leq 1/2$. Indeed, using (8) one has

$$a_+(p) \leq w(p) \leq 4a_+(p), \quad 0 \leq p \leq 1/2,$$

i.e., $a_+(p)$ and $w(p)$ are roughly equivalent. Note that the terms in $R$ have a two-fold representation:

$$w_i(x_i - p_i)(y_i - p_i) = w_i(\bar{x}_i - \bar{p}_i)(\bar{y}_i - \bar{p}_i).$$

Consequently, the $\{w_i\}$ in (7) should be interpreted in the same manner with respect to the prediction of the sequences $\{y_i\}$ and $\{\bar{y}_i\}$.

*2.2. Models of the R-score.*

The weight (9) too heavily emphasizes the complexity of guessing rare events. An alternative is the following parametric family of $w(p)$:

$$w_\beta(p) = [4p(1-p)]^{-\beta}, \quad \beta \geq 0, \tag{10}$$

which includes (9) ($\beta = 1$) and $w(p) = 1$ ($\beta = 0$); in addition, $w_\beta$ possesses the monotone property
$$w_\beta(p) \leq w_{\beta'}(p), \ 0 \leq \beta \leq \beta'.$$
The family is normalized by the requirement $w_\beta(1/2) = 1$.

The weight
$$\tilde{w}_\beta(p) = 1 - \beta \ln[4p(1-p)] \tag{11}$$
for any $\beta \geq 0$ lies between $w_\beta(p)$ and $w_0(p)$, i.e., $w_0 \leq \tilde{w}_\beta \leq w_\beta$.

When $p$ is small, the dominant part of $\tilde{w}_\beta$ is proportional to $\ln(1/p)$, i.e., to the Shannon information, which is carried by an event occurring with probability $p$. This is the reason why (11) may be interesting for the analysis of prediction results as well.

In the GS approach, the complexity measure $a_+(p)$ must decrease on $(0,1)$, while $b_+(p)$ must be a non-decreasing function on $(0,1/2)$. For the family (10) one has
$$a_+(p) = cp^{-\beta}(1-p)^{2-\beta}, \ b_+(p) = c[p(1-p)]^{1-\beta},$$
i.e., the above properties of $a_+$ and $b_+$ are satisfied when $0 \leq \beta \leq 1$. The same statement is true with regard to $\tilde{w}_\beta(p)$.

Note that $R_w$ with $w = w_{1/2}(p)$ and $w = w_1(p)$ admits of an additional statistical treatment. When $\beta = 1/2$, the weights $w_i$ normalize the stochastic terms $(y_i - p_i)$ of $R$ because the quantities
$$w_i(y_i - p_i) = (y_i - p_i)/\sqrt{p_i(1-p_i)} := y_i^{norm}$$
have zero means and unit variances under condition (1). For the same reason, when $\beta = 1$ and $X$, $Y$ have the distribution (1), the score $R(X,Y)$ is the correlation of the normalized vectors $\{x_i^{norm}\}$ and $\{y_i^{norm}\}$, where $x_i^{norm} = (x_i - p_i)/\sqrt{p_i(1-p_i)}$. More generally, by (8) one has
$$R_w(x,y) = \sum_{i \geq 1} b(p_i) x_i^{norm} y_i^{norm}, \ b(p) = b_\pm(p). \tag{12}$$

The representations (7) and (12) are of interest as regards the question of which model of $w(p)$ should be considered "natural". According to the information arguments, this may be $\tilde{w}_\beta(p)$, while the statistical interpretation of (12) leads to the $w_1(p)$ for which $b(p) = 1$ in (12). This indeterminacy cannot be removed without additional argumentation.

The GS approach is far from being the only method for choosing the appropriate measure to assess the performance of a prediction algorithm. As an illustration we consider an example that is conceptually similar to the *entropy score* (see *Vere-Jones, 1998*).

The likelihood of $y_i$, as well as that of $\bar{y}_i = 1 - y_i$, has the form
$$l_i = y_i \log p_i + (1 - y_i) \log(1 - p_i).$$
The quantity $(-l_i)$ has the meaning of the amount of information the observer gets from the i-th prediction experiment. The goal of the forecaster may be formulated as follows: the sum $\sum_i (-l_i)$ should be maximized for those alarm zones where target events are to be expected and minimized where such events are not expected. Hence the goal function may be defined as

$$R_{LH}(X,Y) = \sum_{i \geq 1}(x_i - \bar{x}_i)(-l_i) = \sum_{i \geq 1}(x_i - 1/2)(y_i - 1/2) \cdot 2\ln(p_i^{-1} - 1) + c, \quad (13)$$

where $c$ is independent of $\{y_i\}$. Comparison of (7) with the right-hand side of (13) reveals differences in the centering of the $X$ vector and in the evenness of the weight functions. Nevertheless, both of these goal functions can be used in prediction analysis. To make this clear, consider the case $p_i = p_+ < 1/2$ for all positive alarms and $p_i = p_- < 1/2$ for all negative alarms. Then a structural similarity between $R_{LH}$ and $R_w$ becomes obvious:

$$R_{LH} = v_a^+ \ln\frac{1-p_+}{p_+} - v_a^- \ln\frac{1-p_-}{p_-} + c,$$

and

$$R_w = v_a^+(1-p_+)w(p_+) - v_a^- p_- w(p_-) + c_1,$$

where $v_a^+$ is the number of successful positive alarms and $v_a^-$ that of false negative ones. In the conditional situation with $v_a^+ + v_a^-$ fixed, $R_{LH}$ and $R_w$ are functions of $v_a^+$ only, therefore the analysis of prediction results will rely on the conventional statistic $v_a^+$.

*2.3. R-score and the significance of prediction results.*

Let $\hat{X}$ and $\hat{Y}$ be samples of $X$ and $Y$, respectively. The quantity $R(\hat{X}, \hat{Y})$ can be used to estimate the significance of prediction results. To do this one has to specify the distribution of $Y$. We assume that the components of $Y$ are independent and their distribution follows (1) (*the $H_0$ hypothesis*). The greater the value of $R(X,Y)$, the better is the prediction method. Therefore, the probability

$$\alpha_y = P_y(R(\hat{X}, Y) \geq R(\hat{X}, \hat{Y})), \quad (14)$$

where $P_y$ is the distribution of $Y$ under $H_0$, provides the observed significance level for the prediction results.

Obviously, one has

$$\alpha_y = P\{\xi_R \geq \hat{\xi}_R\},$$

where $\xi_R$ is a linear function of $\{y_i\}$, while $\hat{\xi}_R$ is the observed value of $\xi_R$.

To be more specific,

$$\xi_R = \sum_{i \geq 1} c_i y_i, \quad (15)$$

where $c_i = (\hat{x}_i - p_i)w(p_i)$ for $R = R_w$, and $c_i = (2\hat{x}_i - 1)\ln[(1-p_i)/p_i]$ for $R = R_{LH}$. The significance level $\alpha_y$ can be found from the distribution of $\xi_R$. Under $H_0$ this is the convolution of distributions $\{F_i\}$ of the type:

$$\dot{F}_i(u) = \delta(u)(1-p_i) + \delta(u - c_i)p_i,$$

where $\delta(\cdot)$ is the delta-function. The convolution operation is convenient for numerical computations and for checking the accuracy of the distribution.

If the scatter of the $\{c_i\}$ is not too large (the case of $R_w$ with $p_i > p_0 > 0$), the significance of the $R$-statistic can be roughly inferred from large values of the normalized quantity $\xi_R$:

$$\xi_R^{norm} = (\xi_R - m_R)/\sigma_R, \quad (16)$$

where $m_R$ and $\sigma_R^2$ are the mean and variance of $\xi_R$:

$$m_R = \sum_{i \geq 1} c_i p_i, \quad \sigma_R^2 = \sum_{i \geq 1} c_i^2 p_i (1 - p_i). \tag{17}$$

The estimation of $\alpha_y$, based on $R$ needs two general comments:

– $\alpha_y$, may strongly depend on the choice of the $\{w_i\}$. Therefore a serious argumentation for the choice of $R$ and, in particular, of $\{w_i\}$, is required;

– usually the parameters $\{p_i\}$ for large events are small and their estimates are inaccurate, hence interval estimates of $\alpha_y$, are required. That problem is difficult for analytical solution, because $\{p_i\}$ are involved in the distribution of $\xi_R$ (see (15)) through the distribution of $Y$ and through the weight coefficients $\{c_i\}$ which can have different signs.

**3. The *R*-score and comparison of prediction methods**

*Zechar & Zhuang (2010),* to be referred hereinafter as [ZZ], use the $R$-score to compare a prediction method of interest $X$ with the method of random guessing $Z$. This point should be discussed.

Let $A_i = G_i \times T_i$ be alarms of $X$. If the target events are poissonian, the $p$-parameters of the $H_0$ hypothesis can be specified as follows:

$$p_i = P\{y_i = 1\} = 1 - \exp(-\lambda_i T_i), \tag{18}$$

where $\lambda_i$ is the rate of target events in $G_i$. By definition, the random strategy $Z$ has the distribution $P_z$, which is identical with the distribution of $Y$ given by (18). To compare $X$ with $Z$ under $H_0$, [ZZ] consider the quantity

$$\alpha_z = P_z\{R_w(Z, \hat{Y}) \geq R_w(\hat{X}, \hat{Y})\}. \tag{19}$$

Here, $\alpha_z$ is the measure of all samples of $Z$ having $R$-scores greater than $\hat{X}$. If $\alpha_z$ exceeds a nominal level $\alpha_0$, a practical conclusion may sound as follows: the method $X$ looks no better than random guessing, i.e., $X \prec_R Z$ or that is more correct, the $R$-score does not detect the preference of $X$ compared with random guessing.

$\hat{X}$ and $\hat{Y}$ are interconnected and may come from a short period of monitoring. Therefore, to compare $Z$ and $X$ under $H_0$ we may consider the following natural alternative to $\alpha_z$:

$$\alpha_{zy} = P_{zy}\{R_w(Z, Y) \geq R_w(\hat{X}, \hat{Y})\}, \tag{20}$$

where $P_{zy}$ is the product measure of $P_z$ and $P_y$. From (12) it follows that $R_w(Z, Y)$ under $H_0$ has zero mean and the variance $\sigma_{zy}^2 = \sum_{i \geq 1} b_i^2$. Hence, a rough conclusion like $X \prec_R Z$ may follow from the relation

$$R_w(\hat{X}, \hat{Y}) < k\sigma_{zy},$$

where $k \approx 2$. For the [ZZ] model of $R$ one has $b_i = 1$, i.e., $\sigma_{zy}^2 = N_a$.

It is important, that the quantities $\alpha_y$, $\alpha_z$, and $\alpha_{zy}$ are not necessarily correlated when different models of $R_w$ are considered.

Indeed, let $S$ be a space of vectors $(Z, Y)$ with the measure $P_{zy} = P_z \circ P_y$. Put $R_w(\hat{X}, \hat{Y}) = c$ and denote by $U_c$ the subset of $S$ where $R_w \geq c$. Then $\alpha_{zy}$ is the $P_{zy}$-measure of $S$, while $\alpha_y$ and $\alpha_z$ are the conditional measures of sections of $S$ given by the relations $Z = \hat{X}$ and $Y = \hat{Y}$, respectively. We illustrate our statement by an example of this construction.

Let $S = [0,1]^2$ be a square with the uniform measure; suppose that $U_c$ is a circle of radius $R = 0.25$ with the center $(0.5, 0.5)$, $\hat{X} = 0.5 - R + \varepsilon$, where $\varepsilon$ is small, and $\hat{Y} = 0.5$. Then $\alpha_y \leq \sqrt{2\varepsilon}$, $\alpha_{zy} \approx 0.2$, and $\alpha_z = 0.5$.

Thus in the general situation, *the conclusion like* $X \prec_R Z$ *does not mean that the method* $X$ *is trivial,* because $\alpha_y$ may be small.

[ZZ] apply their method of the comparison to the RTP algorithm (*Shebalin et al., 2004, 2006*). In this application
– the positive RTP alarms $A_i = G_i \times T_i$ are considered only, therefore $\hat{X} = (1,\ldots,1)$;
– the space-time alarm zones of the random strategy $Z = (z_1, \ldots, z_{N_a})$ are the same as the RTP alarms: $T_i \approx 9$ months, the areas $G_i$ are not of regular shape, they represent the union of a standard local areas determined by current and past seismicity, i.e., the alarm zones of $Z$ are fixed and not involved in the simulation. Therefore the solution $z_i = 1$ means that the i-th RTP alarm $A_i$ remains in force, while the solution $z_i = 0$ converts the positive i-th alarm into the negative one.

As a result, $\alpha_z$ shows how efficient is the alarm cancellation mechanism: it is ineffective if $\alpha_z < \alpha_0$ and non-trivial otherwise. According to [ZZ], the RTP alarms admit of different options and interpretations. The so called *loose* interpretation involves 6 successful alarms and gives $\alpha_z \approx 0.0001$ for the $R_w$ model (9). In contrast to this, the more *strict* option involving two successes only gives $\alpha_z \approx 0.94$. These, so opposite, estimates of $\alpha_z$ mean that the question of advisability of random cancellation of RTP alarms remains open. At the same time, $\alpha_z$ gives no information about the significance level, $\alpha_y$, of the RTP alarms.

It is our opinion that the statistical analysis of any prediction method with few target events and a short monitoring period is premature (this is the case of RTP). For this reason we will consider the significance $\alpha_y$ of the M8 algorithm prediction results. For the sake of simplicity we shall use below the notation $\alpha$ for the quantity $\alpha_y$.

## 4. Analysis of the M8: the conventional approach

Description of the M8 prediction algorithm can be found in *Keilis-Borok & Kossobokov (1990)*, *Kossobokov et al. (1990)*, *Kossobokov & Shebalin (2003)*. A statistical analysis of the M8 prediction results was considered recently by *Molchan & Romashkova (2010)* using conventional methods. This enables us to provide below a comparative analysis of the GS and conventional approaches based on a standardized use of earthquake catalogs.

We consider the M8 prediction results for $M \geq 8.0$ events. This prediction has been conducted since 1985, but we will only discuss the period of forward testing of the M8 algorithm, i.e., 1992-2009 (see http://www.mitp.ru/en/predlist.html). The monitoring space $G$ consists of a set of overlapping circles $B_R$ of radius $R = 668$ km located along the Circum-Pacific and Alpine-Himalayan belts. Any circle is permanently in an alarm/non-alarm state. The states are revised for all circles simultaneously at intervals of six months, $\Delta t = 0.5$ year. The union of circles that have been in a state of alarm during six months $\Delta t_i$ will be treated (in the GS approach) as a single domain of positive alarm, $A_i^+ = G_i \times \Delta t_i$; the domain $A_i^- = G_i^c \times \Delta t_i$ where $G_i^c$ is the complement of $G_i$ in $G$ will be considered as a negative alarm.

Table 1 summarizes some results from our analysis of the M8 algorithm for predicting $M \geq 8.0$ events (see *Molchan & Romashkova, 2010)*. Some comment is in order.

The similarity principle, which is the basis of the M8 algorithm, requires specifying the magnitude range of the target $M \geq 8.0$ events. We consider two options: $8.0 \leq M < 8.5$ and $8.0 \leq M < 8.7$; of these two, the former is the more logical, since there is a special version of the M8 algorithm for predicting the $7.5 \leq M < 8.0$ events and this version uses the same similarity principle.

The prediction results for the monitoring period $T = 1992 - 2009$ involve:
- the number of target events $N_e$ and the number of predicted events $v_e^+$. For the option $8.0 \leq M < 8.5$ one has $v_e^+ / N_e = 10/18$;
- a normalized measure of space-time alarm $\tau$:

$$\tau = \lambda(A^+)/\lambda(G \times T), \qquad (21)$$

where $\lambda(A)$ is the expected number of target events in the space-time domain $A$, $A^+$ is the union of all positive alarms $\{A_i^+\}$, and $G \times T$ is the entire prediction space-time volume. Below we discuss two estimates for $\tau$: $\hat{\tau} = 32.5\%$ and $\tilde{\tau} = 35.4\%$.

Because the target events are few, the estimate of $\lambda(\cdot)$ is based on the hypothesis that the $M \geq M_-$ seismicity is stationary and on the regional Gutenberg-Richter (G-R) relations for $M_- \leq M \leq 8.0$ (see for details *Molchan & Romashkova, 2010)*. As a result, we find for $\tau$ a point estimate $\hat{\tau}$ and an upper estimate $\tilde{\tau}$ with confidence level 99%. Both of these estimates are stable with respect to magnitude type ($Mw$ or $Ms$) and to the threshold $M_-$, provided the threshold magnitude is reported completely.

The choice of $M_-$ is influenced by two opposite tendencies: with increasing $M_-$ the G-R law hypothesis becomes more likely; at the same time the amount of available data $N_\lambda = \#\{M \geq M_-\}$ decreases, thereby making the uncertainty of $\tau$ larger. Our analysis (see *Molchan & Romashkova, 2010*) shows that the following restriction on $N_\lambda$ is reasonable:

$$N_\lambda / areaG > 100 / areaB_R, \qquad (22)$$

where $B_R$ is the space unit of M8 alarms, i.e., a circle of radius $R$.

We use $Mw$ as the magnitude most consistent with the G-R law for the range $(M_-, 8.0)$. The resulting estimates $\hat{\tau} = 32.5\%$ and $\tilde{\tau} = 35.4\%$ are obtained with $M_- = 5.5$ and $N_\lambda = 8500$. The data are from the CMT catalog (*Ekstrom et al., 2005*), 1977-2004.

The significance of prediction results is based on the conditional distribution of $v_e^+$ given $N_e$. Under the Poisson hypothesis for the target events, $v_e^+$ has the binomial distribution with parameters $(N_e, \tau)$, which gives the significance level

$$\alpha = P\{v_e^+ \geq \hat{v}_e^+ \mid N_e = \hat{N}_e\},$$

where $\hat{\nu}_e^+$, $\hat{N}_e$ are samples of $\nu_e^+$, $N_e$. The estimate $\tau = \hat{\tau}$ leads to the point estimate $\alpha = \hat{\alpha}$, while $\tilde{\tau}$ leads to the upper estimate, $\alpha \leq \tilde{\alpha}$. For the [8.0, 8.5) option one has $\hat{\alpha} = 3.7\%$ and $\tilde{\alpha} = 6.4\%$.

Since $\hat{\nu}_e^+$ and $\hat{N}_e$ are small for the forward M8 monitoring, the upper estimate $\tilde{\alpha}$ is unstable over time. In fact, a next target event in the monitoring zone will modify the pair $(\hat{\nu}_e^+, \hat{N}_e)$ to become $(\hat{\nu}_e^+, \hat{N}_e + 1)$ or $(\hat{\nu}_e^+ + 1, \hat{N}_e + 1)$. As a result, the estimate $\tilde{\alpha} = 6.4\%$ transforms to become 9.4% or 3.8%, respectively.

**Table 1.** Significance level $\alpha$ for M8 prediction results based on the number of predicted events $\nu_e^+$. *Notation*: $N_e$ is the number of target events; $\hat{\tau}$ and $\tilde{\tau}$ are a point estimate and an upper estimate, respectively, for the normalized measure of space-time alarm $\tau$; $\hat{\alpha}$ and $\tilde{\alpha}$ are a point estimate and an upper estimate of $\alpha$, respectively; $\hat{\alpha}_{PS}$ is a point estimate of $\alpha$ based on $Ms \geq 8.0$ earthquakes from the catalog by *Pacheco and Sykes (1992)*, $\#\{Ms \geq 8.0\} = 63$.

| Period, T | Target events | $\nu_e^+/N_e$ | $\hat{\tau}$ | $\hat{\alpha} \cdot 100\%$ | $\tilde{\tau}$ | $\tilde{\alpha} \cdot 100\%$ | $\hat{\alpha}_{PS} \cdot 100\%$ |
|---|---|---|---|---|---|---|---|
| 1992-2009 | $8.0 \leq M < 8.5$ | 10/18 | 0.325 | 3.7 | 0.354 | 6.4 | 3.2 |
| | $8.0 \leq M < 8.7$ | 11/21 | | 4.7 | | 8.3 | 4.0 |

More information on the issues here discussed can be found in *Molchan & Romashkova (2010)*.

## 5. The *R-s*core in the M8 analysis

*5.1 General remarks*

For the monitoring period $T = 1992 - 2009$ we have $N_+ = 36$ positive alarms $\{A_i^+\}$ and the same number of negative ones $\{A_i^-\}$, each lasting six months. Almost all alarms are not simply connected; this circumstance will be discussed later. A positive alarm $A_i^+$ is treated as successful if it contains at least one target event; a negative alarm $A_i^-$ is treated as successful if it does not contain target events. For this reason in the GS approach the number of successful positive alarms $\nu_a^+$ and the total number of alarms $N_a$ that cover target events are not necessarily equal to the number of predicted events $\nu_e^+$ and the number of all target events $N_e$, respectively. In particular, for the option $8.0 \leq M < 8.5$ we have $\nu_e^+/N_e = 10/18$ and $\nu_a^+/N_a = 10/17$.

The probabilities $p_i^\pm$ for the alarms $A_i^\pm$ are found from (18) using the rates of target events. The method for estimating the rates has been mentioned above. The alarms $A_i^\pm$ are ordered in time, so the plots $i \to p_i^\pm$ are shown as functions of time in Fig. 1.

Histograms for $\{p_i^+\}$ and $\{p_i^-\}$ (Fig. 2a) give an idea of the ranges of $p^\pm$: (0.1, 0.25) for $p^+$ and (0.26, 0.39) for $p^-$, as well as demonstrate typical values of these quantities. It is important for subsequent argument that the $p$-values for positive M8 alarms are bounded away from zero: $p_i^+ \geq 0.1$ for all alarms, and $p_i^+ \geq 0.15$ for successful ones.

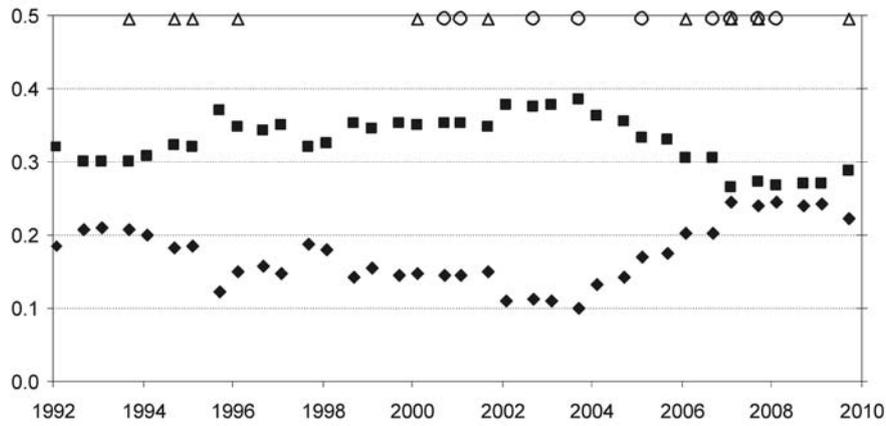

**Figure 1.** The $p$-value for time sequences of six-month M8 alarms ordered in time: positive alarms (*diamonds*), and negative alarms (*squares*). *Notation*: Alarms that have covered target events are marked on the upper horizontal axis: successful positive alarms (*triangles*) and false negative alarms (*circles*).

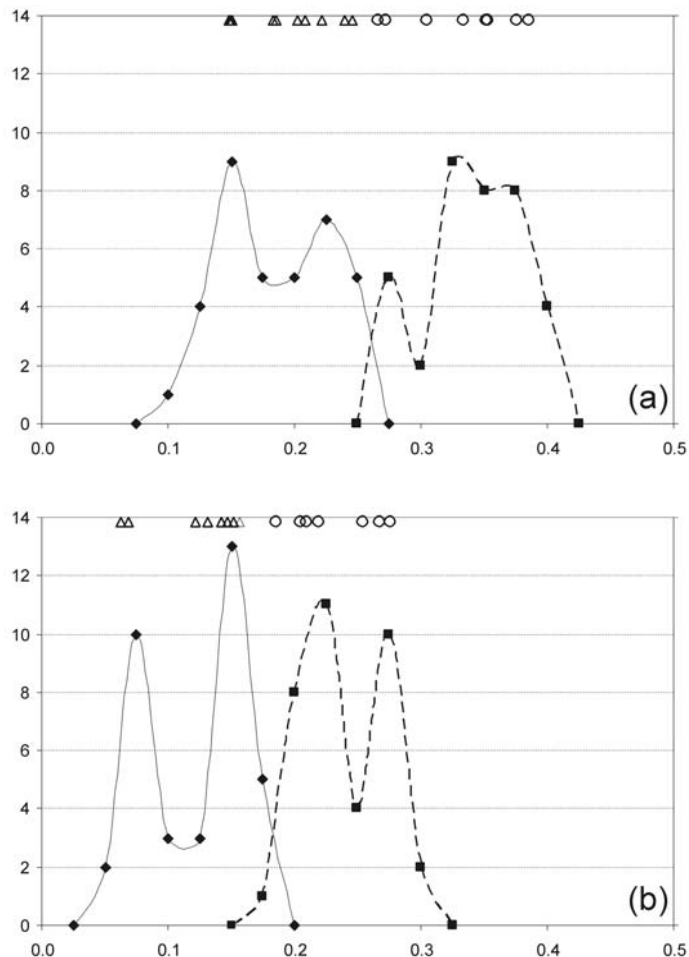

**Figure 2.** The histograms of parameters $\{p_i\}$ for M8 alarms: positive alarms (*diamonds*) and negative alarms (squares). The $p$-values are based on the following data: **(a)** $Mw \geq 5.5$ (see Fig.1), **(b)** $Ms \geq 8.0$ earthquakes from the catalog by *Pacheco and Sykes (1992)*. *Notation*: The upper horizontal axes mark the p-values for the alarms that cover target events: successful positive alarms (*triangles*) and false negative alarms (*circles*).

Below we use the $R$-scores $R_{LH}$ and $R_w$ with four types of $w$:

$$w_0 \leq \tilde{w}_{1/2} \leq w_{1/2} \leq w_1, \tag{23}$$

see (10), (11). According to Section 2.3, the significance level $\alpha$ based on the $R$-score statistic is

$$\alpha = P(\xi_R \geq \hat{\xi}_R), \tag{24}$$

where $\hat{\xi}_R$ is an observed value of $\xi_R$. Now $\xi_R$ is $\xi_R^+ - \xi_R^-$, $\xi_R^+$, or $-\xi_R^-$ if all, positive, or negative alarms are considered, respectively. Here

$$\xi_R^\pm = \sum_{i \geq 1} c^\pm(p_i^\pm) y_i^\pm, \tag{25}$$

where $y_i^\pm = 1$, if $A_i^\pm$ contains a target event and $y_i^\pm = 0$ otherwise. The functions $c^\pm(p)$ are

$$R_w: \quad c^+(p) = w(p)(1-p), \quad c^-(p) = w(p)p \tag{26}$$

and

$$R_{LH}: \quad c^+(p) = c^-(p) = \ln[(1-p)/p]. \tag{27}$$

*5.2 Relationship between $\alpha$ and $w(p)$: a theoretical analysis*

The functions $c^\pm(p)$ for weights (23) are shown in Fig. 3. The $c^-(p)$ vary slowly and are little sensitive to the model of $w$, especially in the range of $\{p_i^-\}$ for the M8 alarms. It follows that the statistic $\xi_{(R,w)}^-$ must be little sensitive to the choice of $w$.

The behavior of $c^+(p)$ is essentially different. Near $p = 0$ we have a rapid decrease of $c^+(p)$ like $O(p^{-\beta})$ for $w = w_\beta$. Far away from $p = 0$ both the decrease of $c^+$ and its dependence on the $w$ model are substantially weaker. Therefore the estimates of $\alpha$ may depend on the behavior of $w$ and on the distribution of $\{p_i^+\}$ in the range (0, 1/2). To illustrate this we consider a model situation.

Suppose the numbers $p_i^+ < p_0$ are such that $c^+(p_i^+)$ are substantially greater than $c^+(p_j^+)$ with $p_j^+ > p_0$. Let us represent the event $\{\xi_{(R,w)}^+ \geq \hat{\xi}_{(R,w)}^+\}$ in the form

$$\sum_{p_i^+ < p_0} c^+(p_i^+) y_i^+ + \Sigma' \geq \sum_{p_i^+ < p_0} c^+(p_i^+) \hat{y}_i^+ + \hat{\Sigma}', \tag{28}$$

where $\Sigma'$ is analogous to the first sum and is applied to $p_i^+ > p_0$, while $\hat{\Sigma}'$ is the observed value of $\Sigma$.

At first we consider the case with $\hat{y}_i^+ = 0$ for $p_i^+ < p_0$ (for M8 this is true for $p_0 = 0.15$). By the assumption $\Sigma'$, $\hat{\Sigma}'$ are substantially smaller than the first sum in (28) and, as we know, are weakly dependent on $w$. The functions $c^+(p)$ increase with $\beta$ for the weights $w = w_\beta(p)$ (see Fig. 3). It follows that the probability of the event (28) must increase with increasing $\beta$, because the first sum increases and dominates in (28). Consequently, the significance of prediction results will be the worst with the model $w = w_1$.

The situation becomes different, if $\hat{y}_{i_0}^+ \neq 0$ for some $p_{i_0}^+ < p_0$. For the sake of simplicity we assume that the sum $\Sigma$ consists of a single term, $i = i_0$. The greater the value of $\beta$ and the smaller the value of $p_{i_0}^+$, the easier one can realize a situation in which $\Sigma'$ and $\hat{\Sigma}'$ will be small compared with the first sum. In that case the probability of (28) will be

$$\alpha \approx P\{c^+(p_{i_0}^+)y_{i_0}^+ \geq c^+(p_{i_0}^+)\} = p_{i_0},$$

that is, when $\beta$ is large, the significance of the $R$-statistic is controlled by a single successful alarm with a very small $p$, and that result may be the best possible.

Since $\xi_{(R,w)}^-$ is weakly dependent on $w$, all considerations outlined above remain valid for the statistic $\xi_{(R,w)}^+ - \xi_{(R,w)}^-$ as well. It remains to verify our heuristic argument by estimating $\alpha$.

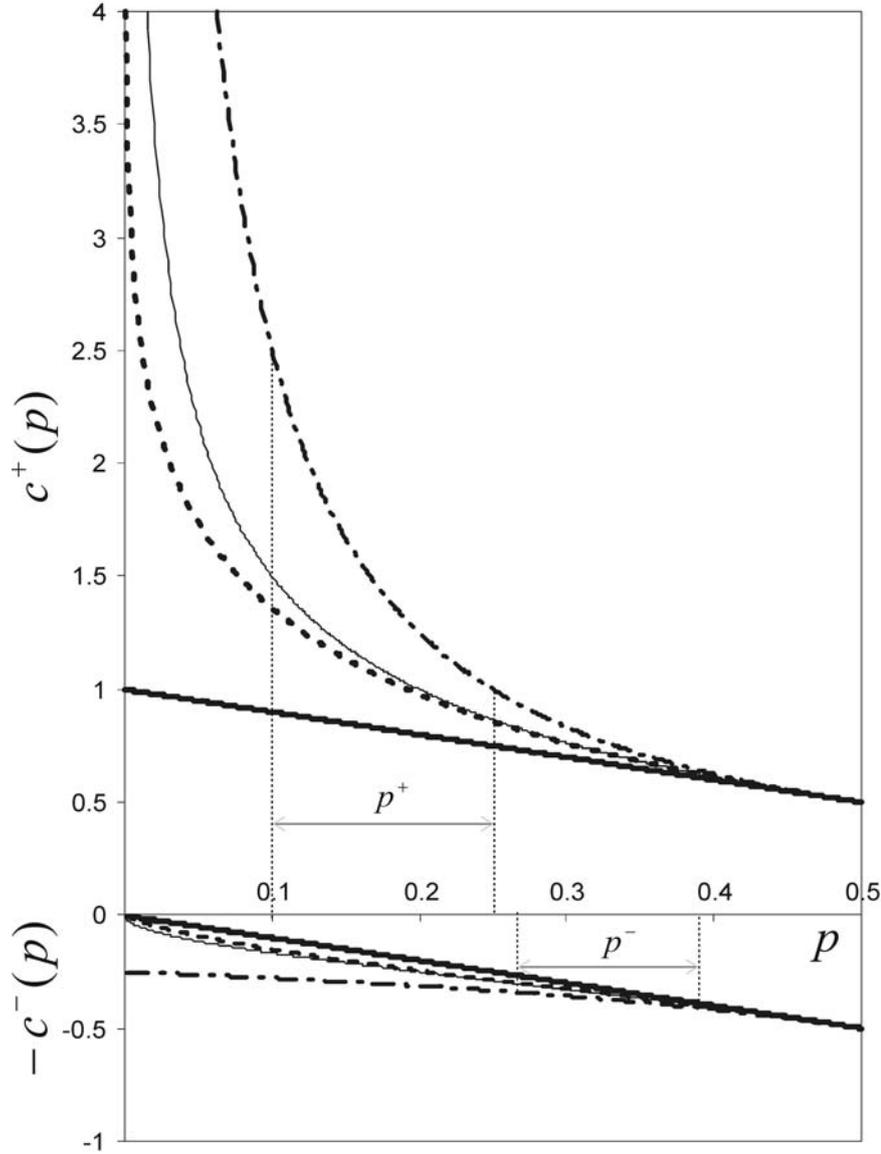

**Figure 3.** Functions $c^+(p)$ (*top*) and $-c^-(p)$ (*bottom*) for different models of the alarm weight function $w(p)$: $w_0$ (*bold line*), $\tilde{w}_{1/2}$ (*dashed line*), $w_{1/2}$ (*thin line*), and $w_1$ (*dash-dotted line*). Vertical dotted lines mark the range of the $p$-value for M8 alarms: negative alarms ($p^-$) and positive alarms ($p^+$).

### 5.3 $\alpha$-estimates for M8 alarms: the GS approach

A rough idea of $\alpha$ is provided by the normalized quantities $\xi_R$, i.e., $\xi_R^{norm}$ (see (16)). They are close in distribution to the standard Gaussian variable, if the number of alarms is large and $\{p_i^+\}$ are bounded away from zero. Therefore the inequality $\xi_R^{norm} > 2$ favors significance of prediction results.

Along with $\xi_R^{norm}$ we give the $\alpha$-estimates based on the exact distribution of $\xi_R$. The uncertainty of $\alpha$ arises due to the fact that the distribution is computed using a discrete grid (the uncertainty increases with increasing number of alarms when a constant grid step is used). The results are summarized in Table 2. They are given for the two magnitude ranges of target events; for three types of alarms: + (positive), - (negative), and +/- (all); and for different $R$-statistics: $R_{LH}$ and $R_w$ with $w$ given in (23).

**Table 2.** Significance level $\alpha$ for M8 prediction results based on the $R$-score: $R_w$ and $R_{LH}$. *Notation*. Alarms: positive (+), negative (-), and all (+/-); $v_a^+/N_a$ is the number of successful positive alarms vs. the total number of alarms that cover target events; $\hat{\alpha}$ is a point estimate of $\alpha$; $\hat{\alpha}_{PS}$ is a point estimate of $\alpha$ based on the $Ms \geq 8.0$ earthquakes from the catalog by *Pacheco and Sykes (1992)*; $\xi_R^{norm}$ is given by (25) in the text.

| $R$-score | $8.0 \leq M < 8.5$, $v_a^+/N_a = 10/17$ | | | | | | | $8.0 \leq M < 8.7$, $v_a^+/N_a = 10/19$ |
|---|---|---|---|---|---|---|---|---|
| | $\xi_R^{norm}$ | | | $\hat{\alpha} \cdot 100\%$ | | $\hat{\alpha}_{PS} \cdot 100\%$ | | $\hat{\alpha} \cdot 100\%$ |
| | (-) | (+) | (+/-) | (+) | (+/-) | (+) | (+/-) | (+/-) |
| $(R, w_0)$ | 1.85 | 1.60 | 2.26 | 7.0 – 7.5 | 1.5 – 1.6 | 0.3 | 0.2 | 3.1 – 3.4 |
| $(R, \widetilde{w}_{1/2})$ | 1.79 | 1.49 | 2.09 | 7.5 – 7.7 | 2.3 – 2.5 | 0.3 | 0.3 | 4.1 – 4.5 |
| $(R, w_{1/2})$ | 1.81 | 1.46 | 2.02 | 7.9 – 8.1 | 2.6 – 2.8 | 0.4 | 0.3 – 0.4 | 4.5 – 4.9 |
| $(R, w_1)$ | 1.78 | 1.29 | 1.75 | 10.3 – 10.7 | 4.6 – 4.9 | 0.9 | 0.8 – 0.9 | 6.7 – 7.2 |
| $R_{LH}$ | 1.59 | 1.74 | 2.35 | 8.6 – 8.9 | 3.2 – 3.5 | 0.4 | 0.6 | 5.7 – 6.1 |

We summarize our conclusions from Table 2:
– like-sign M8 alarms are not significant: for the $8.0 \leq M < 8.5$ option and for any model of $R$, (+) and (-) alarms give $\xi_R^{norm} < 2$; $\alpha = 7 - 11\%$ for the positive M8 alarms. As was to be expected, $\xi_R^{norm}$ for negative alarms is nearly independent of $w(p)$, $\xi_R^{norm} \approx 1.8$;
– for all types of alarm, (+), (-), and (+/-), the significance of prediction results grows as $w_\beta$ varies from $w_1$ to $w_0$, i.e., $\xi_R^{norm}$ increases while $\alpha$ decreases. This is in agreement with our theoretical analysis of the case where the positive alarm with the smallest $p$-value is false (see 5.2);

- of the two $\Delta M$ options, the lowest values of $\alpha$ are obtained for $8.0 \leq M < 8.5$. This fact agrees with Table 1 and supports the preliminary arguments in favor of the [8.0,8.5) option (see 5.1);
- the $\alpha$ for the $8.0 \leq M < 8.5$ option varies between 1.55% ($w = w_0$) and 4.75% ($w = w_1$), thus confirming that the M8 algorithm is non-trivial. It is important that the simple statistic $v_e^+$ gives a similar point estimate, $\alpha = 3.7\%$ (see Table 1).

**Table 3.** Significance level $\alpha$ for M8 alarms depending on the $p$-value for a particular positive alarm, $i_0$: original ($p_{i_0}^+ = 0.15$) and hypothetical ($p_{i_0}^+ = 0.05$). The results are shown for three $R$-models.

| Target events | $p_{i_0}^+$ | $\hat{\alpha} \cdot 100\%$ | | |
|---|---|---|---|---|
| | | $(R, w_0)$ | $R_{LH}$ | $(R, w_1)$ |
| $8.0 \leq M < 8.5$ | 0.15 | 1.55 | 3.35 | 4.75 |
| | 0.05 | 1.20 | 1.8 | 0.95 |
| $8.0 \leq M < 8.7$ | 0.15 | 3.25 | 5.9 | 6.95 |
| | 0.05 | 2.6 | 3.3 | 1.45 |

Note that we have considered one of the possible variants of subdivision of the M8 alarm volume into the isolated parts $A_i^\pm$. To estimate the $\alpha$, a set of simply connected alarms may be more reasonable. However, any additional splitting of the alarms decreases $p_i^+$, hence produces instabilities in the estimation of $\alpha$. The risk involved in the instability of $\alpha$ for the model $w = w_1$ is higher than that for $w = w_0$. To illustrate this we consider the following numerical experiment.

Let $i_0$ be the ordinal number of the M8 positive successful alarm having the smallest $p^+$-value among all successful ones. It is the alarm for the first six months of 2003, $p_{i_0}^+ = 0.15$ (see Fig. 1). Suppose the parameter has been revised so as to make $p_{i_0}^+ = 0.05$. This uniquely specifies $p_{i_0}^-$, since for any $i$

$$(1 - p_i^+)(1 - p_i^-) = \exp(-\Lambda \cdot \Delta),$$

where $\Lambda = \lambda(G)$ is the rate of target events in $G$. Let the other $p$-parameters remain unchanged. Then one has $p_{i_0}^+ = \min_i p_i^+$. Such a case has been considered in Section 5.2. As was to be expected, $\alpha$ is the smallest for the model $w = w_1$ (see Table 3, $p_{i_0}^+ = 0.05$). However, the $\alpha$ based on $w = w_0$ seems more preferable because of the stability, e.g., when we use the [8.0, 8.7) option and $w = w_0$, we have $\alpha = 2.6\%$ (the hypothetical $p_{i_0}^+$) vs. $\alpha = 3.25\%$ (the original $p_{i_0}^+$). Similar $\alpha$-estimates for $w = w_1$ are 1.45% vs. 6.95%, respectively. In other words, in the case $w = w_1$ the original conclusion that the prediction results are not significant becomes its opposite.

*5.4 The R-score with unreliable p-values.*

Our $\{p_i\}$ values for $M \geq 8.0$ are based on the $Mw \geq M_- = 5.5$ earthquakes. This data can include aftershocks in addition to main shocks, hence the estimates of $\lambda_i = \lambda(G_i)$ and $p_i^{\pm}$ may be overestimated. There is a practice of estimating $\{\lambda_i\}$ directly based on past target events, even when $G_i$ contains a single event (see e.g. [ZZ]). The uncertainty of $\alpha$ for $\nu_e^+$ has been investigated in relation to the quantity $N_\lambda = \#\{M \geq M_-\}$ (*Molchan & Romashkova, 2010*). A similar analysis for the $R$-statistic is more difficult. For this reason we repeat the analysis of the M8 predictions using for estimation of $\{p_i\}$ all $Ms \geq 8.0$ earthquakes for the period 1900-1984 from the Global Catalog by *Pacheco and Sykes (1992)* where $N_\lambda = 63$.

Because $N_\lambda$ is small, the estimates of $\{\lambda_i\}$ are highly unreliable. Nevertheless the point estimates of $\tau, \tau_{PS}$, remain practically unchanged; in particular, for the $8.0 \leq M < 8.5$ option one has $\hat{\tau}_{PS} = 0.32$ vs. the original $\hat{\tau} = 0.33$ (Table 1). As a result, the point estimates of $\alpha$ based on $\nu_e^+$ are stable too: $\hat{\alpha}_{PS} = 3.2\%$ vs. $\hat{\alpha} = 3.7\%$ (Table 1). At the same time, the $\alpha$-estimates based on the $R_w$ statistics have diminished considerably: for the $8.0 \leq M < 8.5$ option one has $\hat{\alpha}_{PS} = 0.2\%$ vs. $\hat{\alpha} = 1.55\%$ (Table 2) with $w = w_0$, and $\hat{\alpha}_{PS} = 0.85\%$ vs. $\hat{\alpha} = 4.75\%$ (Table 2) with $w = w_1$. Even when positive alarms only are considered, the new $\alpha$-estimates are rather optimistic, e.g., $\hat{\alpha}_{PS} = 0.9\%$ vs. $\hat{\alpha} = 10.5\%$ (Table 2) with $w = w_1$ and the $8.0 \leq M < 8.5$ option.

These results reveal a fundamental difference between the conventional and the GR approach. The first operates with the conditional distribution of $\nu_e^+$ and hence with the relative characteristic $\tau$ (see (21)). In other words, a rescaling of the rate parameters, $\lambda_i \to \rho\lambda_i$, does not affect $\tau$. This transformation of $\{\lambda_i\}$ arises when the magnitude range of target events, $M \geq M_0$, is shifted by $\delta M \approx -\lg \rho$. If $M_0$ is fixed, the shift may have been caused by the fact that a different catalog or a different magnitude type is used. That is why the $\tau$-estimates for the M8 alarms weakly depend on the choice of the $M$-scale and on the magnitude range $M > M_-$ used to estimate $\{\lambda_i\}$ (see *Kossobokov, 2005; Molchan & Romashkova, 2010*).

In contrast to this, the GR approach uses absolute values of $\{\lambda_i\}$. Therefore, the estimates of $\alpha$ may be sensitive to the transformation $\lambda_i \to \rho\lambda_i$. Figure 2b shows the histogram of $\{p_i\}$-estimates based on $Ms \geq 8.0$ events from the *Pacheco & Sykes (1992)* catalog. We can see that the range of $\{p_i^+\}$ is (0.05, 0.2) (Fig. 2b) vs. the previous (0.1, 0.25) (Fig. 2a). As a result, we have more optimistic estimates of $\alpha$, $\hat{\alpha}_{PS}$.

## 6. Conclusion and Discussion

- The Gambling Score approach takes into account both the positive or/and negative alarms and the probability, $p$, of an expected event in each alarm zone. To estimate the forecaster's skill in the best way each result of prediction is considered with a weight that depends on $p$. A fair scoring scheme helps to reduce (but not entirely remove) the number of unknown parameters in the gambling score $R$ (see (7) for the arbitrary weight function $w(p)$ in $R$). For this reason a serious argumentation for the choice of the $R$ model is required in every case of its application.

- Basing on $R$-score *Zechar & Zhuang (2010)* suggest a method for comparison of predictions with the random guess. Theoretical arguments show that any result of the comparison cannot exclude the possibility for an observed value $R$ to be significant relative to the reference model of seismicity (non-triviality of the predictions).
- The problem of predictability of strong earthquakes is still debatable. For this reason we apply the GS approach to the significance analysis of results from prediction of $M \geq 8.0$ events by the M8 algorithm.

Theoretical considerations and straightforward estimates of the significance level $\alpha$ show a strong dependence of $\alpha$ on the behavior of the weight function $w(p)$ and on the distribution of alarm $p$-values near zero. Both of these factors may affect the estimate of $\alpha$ in either direction and it can be exploited by the researcher. At the same time, the distribution of the $p$-parameters is affected by:

– the partitioning of the entire alarm space-time into sub-areas, that is, into individual alarms $A_i$;

– the method used to estimate the rate of target events in the alarms, $\{\lambda_i\}$; this involves the choice of the catalog and the $M$-scale, the choice of small events to extrapolate the recurrence of target events, and so on.

The least stable estimates of $\alpha$ result from the weight $w = w_1$ proposed by *Zechar & Zhuang (2010)*.

- All estimates of $\alpha$ based on the $R$ models provide evidence of the M8 algorithm being nontrivial for the prediction of $8.0 \leq M < 8.5$ events: $1.5\% < \alpha < 4.7\%$. This interval covers the estimate $\alpha = 3.7\%$ which was obtained in the conventional way using the number of predicted events, $v_e^+$.

- The estimation of $\alpha$ based on the statistic $v_e^+$ has several important advantages:

– there are analytical upper bounds for $\alpha$ that incorporate the number of data $N_\lambda$ used for estimating the rate of target events $\{\lambda_i\}$;

– estimates of $\alpha$ are not affected when $\{\lambda_i\}$ are replaced with $\{\rho\lambda_i\}$. This makes $\alpha$ stable under the choice of the estimation method for $\{\lambda_i\}$;

– estimates of $\alpha$ do not require any partitioning of the alarm space. The more detailed is the partitioning, the more difficult it is to interpret $\alpha$. For small alarm domains $\alpha$ may merely reflect the uncertainty in $\{\lambda_i\}$.

- The following quantities are important in the statistical analysis of a prediction method: the number of target events $N_e$ for the monitoring period, the number of events $N_\lambda$ used to estimate $\{\lambda_i\}$, and the rate of growth of $w(p)$ near $p = 0$ in case the GR approach is used. Small values of $N_e$, $N_\lambda$ and high rates of growth of $w(p)$ near $p = 0$ destabilize the estimates of significance of prediction results. For this reason the rules that regulate the testing of prediction algorithms should include restrictions on the above quantities. In our analysis of the M8 algorithm we have $N_e \approx 20$, the restrictions (22) for $N_\lambda$, and we consider the statistics: ($R_w, w(p) = 1$) and $\left(v_e^+, \tau\right)$ as the most preferable.